\documentclass[aps,prl,twocolumn,superscriptaddress,english]{revtex4}
\usepackage{amssymb}
\usepackage{graphicx}
\usepackage{amsmath}
\usepackage{soul}
\usepackage{amsthm}
\usepackage{amsfonts}
\usepackage[T1]{fontenc}
\usepackage[latin9]{inputenc}
\usepackage{array}
\usepackage{multirow}
\usepackage{color}
\usepackage{esint}
\usepackage{bm}
\usepackage{color}
\usepackage{bbm}
\usepackage{hyperref}
\usepackage{babel}
\usepackage{titlesec}

\makeatletter
\def\iddots{\mathinner{\mkern1mu\raise\p@
		\hbox{.}\mkern2mu\raise4\p@\hbox{.}\mkern2mu
		\raise7\p@\vbox{\kern7\p@\hbox{.}}\mkern1mu}}
\def\adots{\mathinner{\mkern2mu\raise\p@\hbox{.} 
		\mkern2mu\raise4\p@\hbox{.}\mkern1mu
		\raise7\p@\vbox{\kern7\p@\hbox{.}}\mkern1mu}}
\makeatother

\begin{document}
\global\long\def\id{\mathbbm{1}}
\global\long\def\ui{\mathbbm{i}}
\global\long\def\ud{\mathrm{d}}

\title{Emergent Dynamical Translational Symmetry Breaking as an Order Principle for Localization and Topological Transitions}  

\author{Yucheng Wang}
\email{wangyc3@sustech.edu.cn}
\affiliation{Shenzhen Institute for Quantum Science and Engineering, Southern University of Science and Technology,
Shenzhen 518055, China}
\affiliation{International Quantum Academy, Shenzhen 518048, China}
\affiliation{Guangdong Provincial Key Laboratory of Quantum Science and Engineering, Southern University of Science and Technology, Shenzhen 518055, China}

\begin{abstract}
Localization transitions constitute a fundamental class of continuous phase transitions, yet they occur without any accompanying symmetry breaking. We resolve this by introducing dynamical translational symmetry (DTS), defined not by the Hamiltonian but by the long-time behavior of local observables. Its order parameter, the time-averaged local translational contrast (TLTC), quantitatively diagnoses whether evolution restores or breaks translational equivalence. We show that the TLTC universally captures the Anderson localization transition, the many-body localization transition, and topological phase transitions, revealing that these disparate phenomena are unified by the emergent breaking of DTS. Our work establishes a unified dynamical-symmetry framework for localization and topological  transitions beyond the equilibrium paradigm, opening new avenues for characterizing nonequilibrium phases.
\end{abstract}
\maketitle


{\em Introduction.---} Understanding the nature of phase transitions has long been a central theme in condensed matter physics.  
Traditionally, the Landau paradigm characterizes continuous phase transitions in terms of spontaneous symmetry breaking and the emergence of local order parameters~\cite{Landau1,Landau2}.  However, several important classes of continuous transitions lie beyond this conventional framework, as they do not involve any symmetry breaking. Among the most prominent examples are topological phase transitions, where distinct phases are distinguished not by symmetry but by topological invariants and boundary modes~\cite{TP1,TP2,TP3,TP4}. The deep understanding of such transitions developed over the past few decades has profoundly expanded our view of quantum matter. In parallel, localization transitions pose an equally profound challenge to the Landau framework.
These include (i) the Anderson localization transition, where single-particle eigenstates evolve continuously from extended to localized as disorder increases~\cite{AL1,AL2,AL3,AL4}, and (ii) the many-body localization (MBL) transition, where interacting disordered systems exhibit a breakdown of ergodicity and thermalization~\cite{MBL1,MBL2,MBL3,MBL4,MBL5}.
Both represent continuous phase transitions that occur without any accompanying symmetry breaking.

The essential distinction between extended and localized phases is encoded in their long-time dynamical behavior. Extended or thermalizing systems evolve toward spatially uniform local observables, whereas localized dynamics preserve spatial inhomogeneity and retain memory of the initial state. This contrast suggests that the relevant symmetry for characterizing localization physics is not the microscopic symmetry of the Hamiltonian, but the spatial structure generated by long-time evolution. Motivated by this, we introduce the concept of dynamical translational symmetry (DTS), which describes whether the dynamics restore translational equivalence among lattice sites. 
Its breakdown signals the failure of dynamical homogenization, thereby distinguishing localized and extended regimes. Remarkably, we further find that DTS breaking can also characterize topological phase transitions.
  
In this work, we establish DTS breaking as a dynamical order principle for localization and topological transitions.  
Preservation of DTS corresponds to ergodic or extended dynamics, while its breaking signals localization, memory retention, or boundary confinement. Through the quantitative framework of the time-averaged local translational contrast (TLTC), we demonstrate that Anderson localization, MBL, and topological phase transitions can all be consistently interpreted as distinct manifestations of the same underlying dynamical symmetry principle, thereby extending the Landau paradigm of symmetry breaking to localization and topological physics.

{\em Dynamical Translational Symmetry.---} 
For a static lattice Hamiltonian $H$,  ordinary translational symmetry is defined by $[T_a,H]=0$, where $T_a$ translates all sites by $a$. This implies $[U(t),T_a]=0$ for the evolution operator $U(t)=e^{-iHt}$. This 
operator-level constraint does not determine the symmetry of long-time local 
observables generated from a localized physical initial state. A typical example is provided by flat-band systems: although the Hamiltonian possesses lattice-translation symmetry and the flat-band eigenstates can be combined into translationally invariant Bloch states, the real-space localization inherent to the flat band constrains wave-packet motion, and the long-time evolution of a localized initial state therefore lacks translational invariance.
By contrast, in disordered or quasiperiodic systems, translational symmetry is explicitly broken at the Hamiltonian level ($[T_a,H]\neq 0$). Yet, in the extended phase, the dynamics are ballistic and essentially identical to those in a periodic lattice. Over long times, local observables become nearly uniform, reflecting the dynamical emergence of translational invariance. As disorder or quasiperiodicity increases and the system undergoes a localization transition, spreading is suppressed, memory of the initial state is retained, and persistent spatial inhomogeneity appears. Traditionally, localization transitions have been regarded as continuous phase transitions, but they are not considered to involve any form of conventional symmetry breaking. These observations inspire a unified dynamical description of localization phases and transitions, and motivate identifying the associated symmetry breaking from the perspective of long-time dynamics.


To quantify the degree of DTS breaking, we introduce the TLTC, defined as
\begin{widetext}
	\begin{eqnarray}
	\label{eq:TLTC_def}
	\mathcal{C}_a^{(O)}(T_f,T_i,j)
	= \frac{1}{T_f-T_i}
	\int_{T_i}^{T_f}
	\Big|\big\langle\psi(0)\big|
	\Big(U^\dagger(t)\,O_j\,U(t)
	- \mathcal{T}_a[U^\dagger(t)\,O_j\,U(t)]\Big)\big|\psi(0)\big\rangle\Big|^2 dt ,
\end{eqnarray}
\end{widetext}
where \(O_j\) is a local observable at site \(j\),
\(|\psi(0)\rangle\) is a normalized initial state (typically a localized wave packet), and
\(\mathcal{T}_a[X]\equiv T_a^\dagger X T_a\). 
The parameters $T_i$ and $T_f$ denote the lower and upper limits of the time window used for averaging. 

The TLTC measures the time-averaged deviation between the local expectation value of an observable and that of its translated counterpart during the dynamical evolution. For an arbitrary initial state \(|\psi_0\rangle\), the long-time average of a local observable $O_j$ is defined as 
\[
\overline{O_j}
=\frac{1}{T}\int_{0}^{T}
\langle\psi(0)|
U^\dagger(t)O_jU(t)
|\psi(0)\rangle\,dt,
\]
If the TLTC vanishes for all $j$ in the long-time limit,
$\lim_{T_f\to\infty}\mathcal{C}_a^{(O)}(T_f,T_i,j)=0$,
then  $\overline{O_j} = \overline{O_{j+a}}$, indicating translational equivalence of long-time averaged local observables and the emergence of spatial homogeneity. 
Conversely, if \(\mathcal{C}_a^{(O)}\) approaches a finite value for some site $j$ such that   
\(\overline{O_j}\ne \overline{O_{j+a}}\),
the dynamics remain spatially inhomogeneous, signaling broken DTS. 
As proved in the End Matter, the TLTC vanishes in the long-time limit in the extended (thermalizing) phase. This implies that translational equivalence of local observables emerges dynamically, even when the Hamiltonian lacks translation symmetry due to disorder or quasiperiodicity.

The TLTC thus functions as a dynamical order parameter: it vanishes in the extended (ergodic) phase
and becomes finite in localized 
or dynamically confined regimes. 
This behavior directly parallels the role of static order parameters in Landau's paradigm.
Importantly, the TLTC is basis-independent and
universally applicable to both single-particle and  interacting systems.
As we shall demonstrate below, the TLTC provides a unifying language for describing
Anderson localization, MBL, and topological transitions.

In evaluating Eq.~(\ref{eq:TLTC_def}), the choice of the averaging window \((T_i, T_f)\)
 is flexible.
 For finite systems, it is not necessary to take the strict limit \(T_f \to \infty\);
 a sufficiently long but finite evolution time already captures the asymptotic behavior.
 For analytical and numerical convenience,
 we set \(T_i = 0\) and \(T_f = T\).  
 It is also unnecessary to compute the TLTC 
 for every lattice site.  
 Since DTS concerns translational equivalence between sites, 
 monitoring a representative site is sufficient, 
 typically the one where the local excitation or contrast
 is maximal at \(t=0\).
 For example, in the study of Anderson localization,
 we consider an initial wave packet \(|\psi(0)\rangle\)
 localized at site \(i_0\),
 and evaluate the corresponding \(\mathcal{C}_a^{(O)}(T, 0, i_0)\).
 If this quantity tends to zero in the long-time limit,
 \(\mathcal{C}_a^{(O)}(T, 0, j)\to 0\) for all \(j\),
 signifying complete restoration of DTS.
Without loss of generality, we set \(a=1\) in the following and denote \(\mathcal{C}_a^{(O)}(T_f, T_i, j)\) simply as \(\mathcal{C}_a^{(O)}(T)\).

{\em Anderson localization transition.---} We first illustrate that DTS breaking can characterize the Anderson localization transition using the Aubry-Andr\'e (AA) model~\cite{AA}, a paradigmatic realization of localization transition in quasiperiodic systems.  The single-particle Hamiltonian is
\begin{equation}
	H_{\text{AA}}
	= -J \sum_{i=1}^{L-1} (c_i^\dagger c_{i+1} + \mathrm{h.c.})
	+ \lambda \sum_{i=1}^L \cos(2\pi\beta i + \phi)\, n_i,
	\label{eq:H_AA}
\end{equation}
where $c_i^{\dagger}$ ($c_i$) creates (annihilates) a particle on site $i$, 
$J$ is the hopping amplitude, $\lambda$ controls the strength of the quasiperiodic potential, $\beta$ is an irrational number, and $\phi$ is the initial phase.  
This model exhibits the localization transition at $\lambda_c / J = 2$:  
for $\lambda < \lambda_c$, all eigenstates are extended, while for $\lambda > \lambda_c$ they are exponentially localized.
Without loss of generality, in the following calculations we fix 
$\beta = (\sqrt{5}-1)/2$ and $\phi = 0$.

To probe DTS, we compute the TLTC,
\begin{equation}
	\mathcal{C}_a^{(P)}(T)
	= \frac{1}{T}
	\int_0^T
	\big|
	P_{i_0}(t) - P_{i_0+a}(t)
	\big|^2\, dt,
	\label{eq:TLTC_AA}
\end{equation}
where $P_i(t)=|\psi_i(t)|^2$ is the instantaneous local probability, and $i_0$ denotes the initial position of the wave packet.  
A vanishing $\mathcal{C}_a^{(P)}$ indicates dynamical restoration of translational equivalence, whereas a finite value signals DTS breaking.

\begin{figure}[t]
	\centering
	\includegraphics[width=1\linewidth]{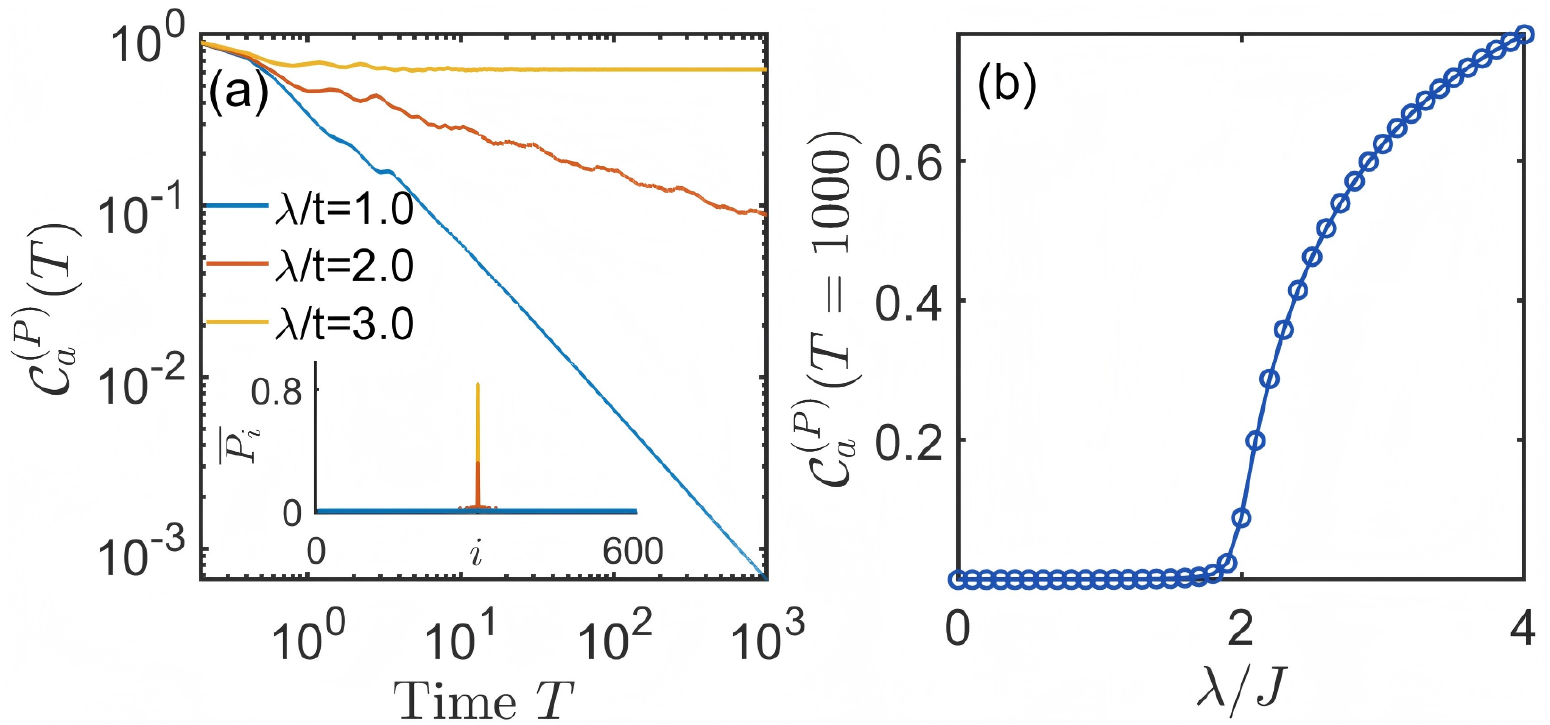}
	\caption{
		(a) Time evolution of the TLTC indicator
		$\mathcal{C}_a^{(P)}(T)$ for representative $\lambda/J=1,2,3$.
		The inset shows the corresponding long-time averaged site probabilities
		$\overline{P_i}$.
		(b) Long-time value $\mathcal{C}_a^{(P)}(T=1000)$ versus $\lambda/J$. Here we fix $J=1$, $L=610$, $i_0=L/2$, $a=1$, use a time integration step of $dt=0.2$, and employ open boundary conditions (OBC).}
	\label{LATC_AA}
\end{figure}

Figure~\ref{LATC_AA}(a) shows the time evolution of $\mathcal{C}_a^{(P)}(T)$ for representative values of  $\lambda/J$.  
In the extended phase ($\lambda/J=1$), $\mathcal{C}_a^{(P)}(T)$ decays algebraically toward zero.    
At the critical point ($\lambda/J=2$), the decay becomes slow and nonmonotonic, reflecting critical fluctuations in the spreading process.    
In the localized phase ($\lambda/J=3$), it rapidly saturates to a finite value.  The inset shows the long-time averaged probability distribution
$\overline{P_i} = \frac{1}{T}\int_0^T P_i(t)\,dt$,
which is uniform in the extended phase and exponentially localized in the localized phase. We fix $T=1000$ and plot the long-time value $\mathcal{C}_a^{(P)}(T=1000)$ as a function of $\lambda/J$ in 
Fig.~\ref{LATC_AA}(b), which shows that $\mathcal{C}_a^{(P)}(T=1000)$ changes from zero to finite across the localization transition. The TLTC thus serves as a dynamical order parameter for the Anderson localization transition by capturing the restoration or breakdown of DTS.

\begin{figure}[t]
	\centering
	\includegraphics[width=1\linewidth]{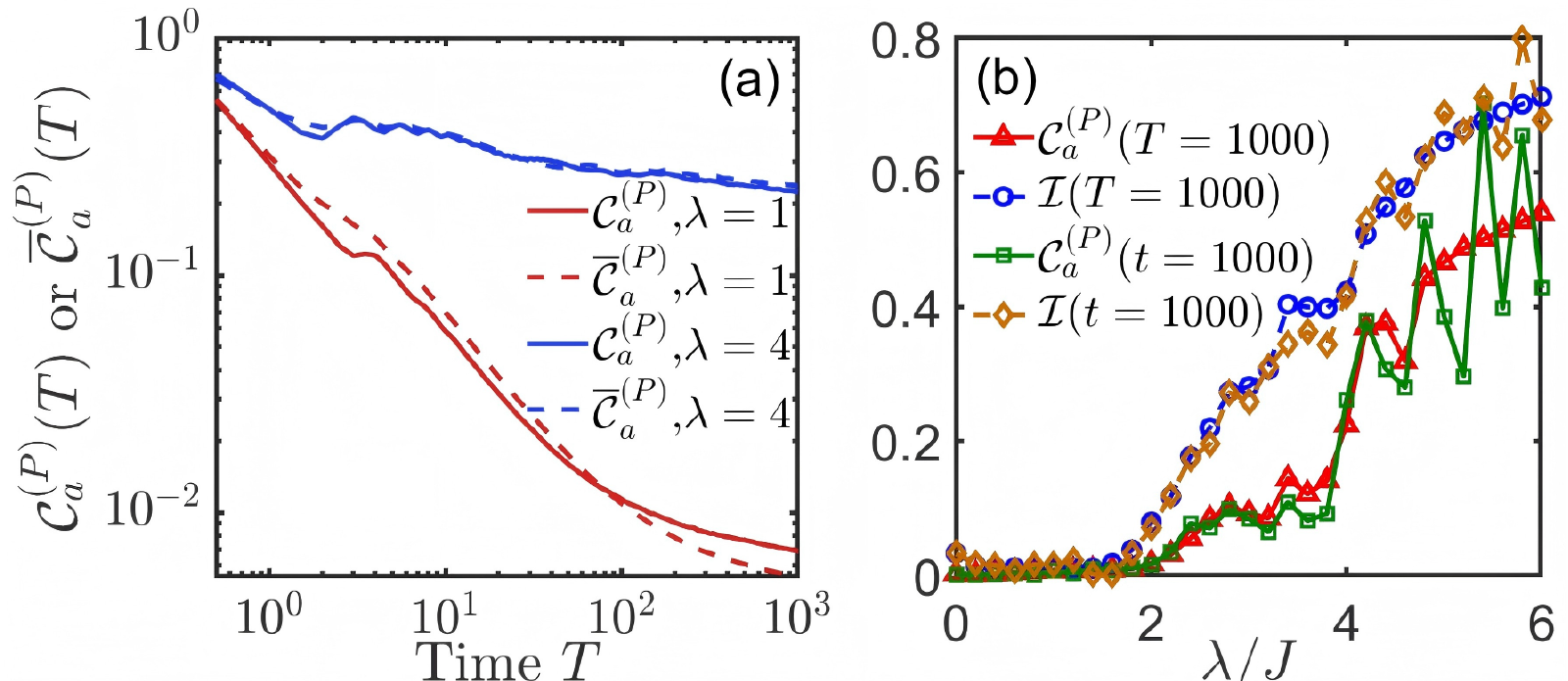}
	\caption{ (a) Time evolution of the TLTC  \(\mathcal{C}_a^{(P)}(T)\) (solid) and its site-averaged counterpart \(\overline{\mathcal{C}}_a^{(P)}(T)\) (dashed) in the interacting AA model at \(\lambda/J=1\) (red) and \(\lambda/J=4\) (blue). 
		(b)~Comparison between the instantaneous and time-averaged TLTC, 
		\(\mathcal{C}_a^{(P)}(t)\) and \(\mathcal{C}_a^{(P)}(T)\), 
		and the density imbalance, 
		\(\mathcal{I}(t)\) and \(\mathcal{I}(T)\), as functions of the quasiperiodic potential strength~\(\lambda/J\). 
		All four quantities change from vanishing to finite values at approximately the same critical~\(\lambda/J\). Here we fix $J=1$, $V=1$, $L=14$, $N=7$, $i_0=7$, $a=1$, $dt=0.5$, and use OBC.
	}
	\label{fig:LATC_AA_MBL}
\end{figure}

{\em Many-body localization transition.---} 
The framework of DTS breaking extends naturally to interacting systems. We consider the interacting AA model,
\begin{equation}
	H_{\text{AA+int}} = H_{\text{AA}}
	+ V \sum_{i=1}^{L-1} n_i n_{i+1},
	\label{eq:H_AA_MBL}
\end{equation}
where \(V\) is the nearest-neighbor interaction strength. The system undergoes an ergodic-to-MBL transition as $\lambda/J$ increases for a fixed $V$~\cite{AAinter}.

We fix the system size to \(L=14\) and the particle number to \(N=7\).  
The initial state is chosen as a charge-density-wave configuration, 
where particles occupy odd lattice sites.  
We employ Eq.~(3) to characterize the distinct dynamical features of different many-body phases, 
with \(P_i(t)=\langle\Psi(t)|n_i|\Psi(t)\rangle\), 
where \(|\Psi(t)\rangle\) denotes the many-body wavefunction at time \(t\).  
Again taking \(a=1\), the reference site \(i_0\) can be chosen arbitrarily, since all occupied and empty sites initially carry maximal contrast.  
For comparison, we also introduce a site-averaged TLTC:  
\begin{equation} 
	\overline{\mathcal{C}}_a^{(P)}(T)
	= \frac{1}{TL}\int_0^T\!\!dt\, 
	\sum_{i=1}^L 
	\big| P_i(t) - P_{(i+a)\,\mathrm{mod}\,L}(t) \big|^2,
	\label{eq:LATC_avg}
\end{equation}
Figure~\ref{fig:LATC_AA_MBL}(a) shows the evolution of
\(\mathcal{C}_a^{(P)}(T)\) and \(\overline{\mathcal{C}}_a^{(P)}(T)\).  
For \(\lambda/J=1\), both quantities decay to zero, reflecting homogeneous densities in the ergodic phase. For \(\lambda/J=4\), they decay slowly and saturate to finite values, signaling MBL and the breaking of DTS.  The two quantities coincide throughout the evolution, confirming that the localization properties can already be faithfully inferred from two sites.

To benchmark against the standard experimental diagnostic of many-body localization,  
we compare the instantaneous and time-averaged TLTC,  
\(\mathcal{C}_a^{(P)}(t)=|P_{i_0}(t)-P_{i_0+a}(t)|^2\) and \(\mathcal{C}_a^{(P)}(T)\) (defined in Eq.~(3)),  
with the experimentally measurable density imbalance \(\mathcal{I}(t)\) and its time average \(\mathcal{I}(T)\), defined as~\cite{MBL4}   
\begin{equation}
	\mathcal{I}(t) = \frac{1}{N}\sum_{i=1}^{L}(-1)^i\, \langle n_i(t)\rangle, 
	\quad 
	\mathcal{I}(T) 
	= \frac{1}{T}\int_0^T \mathcal{I}(t)\, dt.
	\label{eq:imbalance_def} 
\end{equation}
A vanishing imbalance, \(\mathcal{I}(t)\!\to\!0\), indicates ergodic dynamics,  
whereas a finite saturation value signals MBL.  
Figure~\ref{fig:LATC_AA_MBL}(b) shows the variation of these quantities as a function of the quasiperiodic potential strength~\(\lambda\).  
One observes that \(\mathcal{C}_a^{(P)}(t)\), \(\mathcal{C}_a^{(P)}(T)\), \(\mathcal{I}(t)\), and \(\mathcal{I}(T)\)  
all transition from vanishing to finite values at approximately the same critical~\(\lambda/J\),  
demonstrating that the TLTC faithfully captures the ergodic-to-MBL transition.  
Unlike the imbalance, however, measuring \(\mathcal{C}_a^{(P)}\) requires monitoring only two lattice sites  
with initially contrasting occupations, such as one occupied and one empty site, regardless of the detailed form of the prepared initial state.  
This simplicity makes TLTC-based characterization particularly favorable for experimental studies of MBL transitions.
The TLTC thus acts as a unified dynamical order parameter for both Anderson and MBL transitions, linking ergodic spreading to localized memory retention through the restoration or breakdown of DTS.

{\em Topological transition.---} The concept of DTS breaking 
can be further extended to describe topological phase transitions, 
where boundary modes intrinsically violate DTS.  
We illustrate this connection using the Su-Schrieffer-Heeger (SSH) model~\cite{SSH}, 
which is given by  
\begin{equation}
	H_{\text{SSH}}
	= -\sum_{i=1}^{L-1}
	\big[J_1\, c_{2i-1}^\dagger c_{2i}
	+ J_2\, c_{2i}^\dagger c_{2i+1} + \mathrm{h.c.}\big],
	\label{eq:H_SSH}
\end{equation}
where $J_1$ and $J_2$ denote alternating hopping amplitudes.
The system is topologically nontrivial for $J_2/J_1>1$
and trivial for $J_2/J_1<1$.
In the topological regime, open boundaries support exponentially localized
zero-energy edge states.

\begin{figure}[t]
	\centering
	\includegraphics[width=1\linewidth]{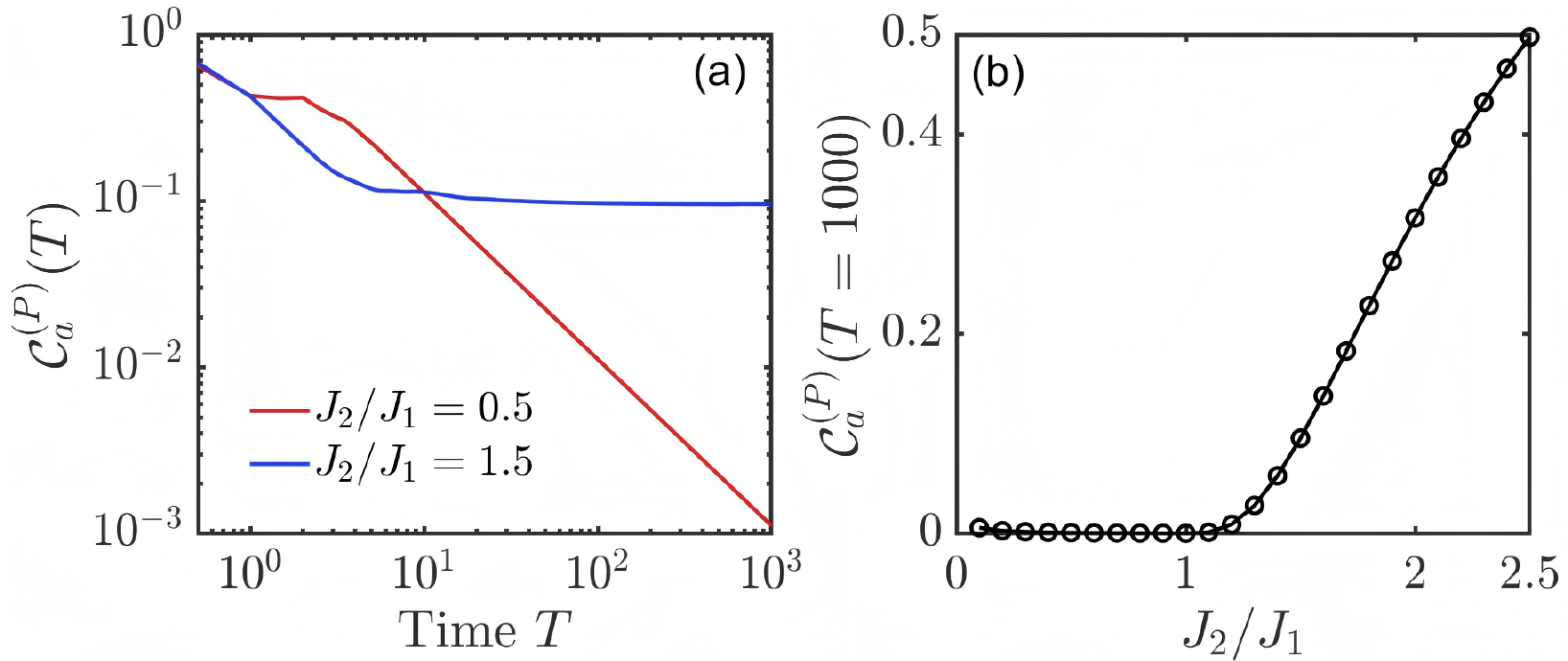}
	\caption{
		(a) Time evolution of the TLTC $\mathcal{C}_a^{(P)}(T)$ in the SSH model for $J_2/J_1=0.5$ (red) and $J_2/J_1=1.5$ (blue), with the particle initially localized at the boundary site.
		(b) The long-time value $\mathcal{C}_a^{(P)}(T{=}1000)$ as a function of $J_2/J_1$.  
		 Here we fix $J_1=1$, $L=600$, $i_0=1$, $a=1$, $dt=0.5$, and use OBC.
	}
	\label{LATCSSH}
\end{figure}

We initialize a single-particle wave packet localized at the boundary site, 
\(\psi_i(0)=\delta_{i,1}\), and monitor its time evolution.  
We again evaluate the TLTC defined in Eq.~(3), taking \(i_0=1\) and \(a=1\).  
As shown in Fig.~\ref{LATCSSH}(a), in the trivial phase (\(J_2/J_1=0.5\)),  
the TLTC \(\mathcal{C}_a^{(P)}(T)\) rapidly decays to zero, 
reflecting that the boundary excitation spreads across the lattice and DTS is dynamically restored.
In contrast, in the topological phase (\(J_2/J_1=1.5\)),  
\(\mathcal{C}_a^{(P)}(T)\) saturates to a finite nonzero value, 
signaling boundary-induced DTS breaking due to the confinement of the edge-localized mode. We fix \(T=1000\), and Fig.~\ref{LATCSSH}(b) shows 
\(\mathcal{C}_a^{(P)}(T=1000)\) as a function of the hopping ratio \(J_2/J_1\).  
As the system evolves from the topologically trivial to the nontrivial regime,  
\(\mathcal{C}_a^{(P)}(T=1000)\) increases from zero to a finite value.  
This demonstrates that the TLTC serves as an effective order parameter 
for the topological transition, directly diagnosing DTS breaking associated with protected boundary states.

{\em Conclusion and discussion.---} We have introduced dynamical translational symmetry (DTS) and formulated the time-averaged local translational contrast (TLTC) as a quantitative measure of its emergence or breaking. 
The preservation of DTS corresponds to ergodic or extended dynamics,
while its breaking signifies localization, memory retention, or boundary confinement.
Although the microscopic origins differ between disorder-induced localized states and topological edge modes, 
their dynamical manifestations share a common structure: both prevent the long-time restoration of translational equivalence.
The TLTC serves as a dynamical order parameter,
analogous to static order parameters in Landau theory,
but defined through the long-time evolution of local observables. Because the TLTC depends only on local observables and time averaging,
it is readily accessible in various experimental platforms,
including ultracold atoms, superconducting quantum circuits, and photonic simulators.

The DTS framework naturally extends to driven and open quantum systems~\cite{explainopen},
where emergent dynamical symmetry may interplay with Floquet synchronization
or dissipation-induced ordering. 
We anticipate that DTS and its breaking will provide
a versatile framework for characterizing nonequilibrium quantum phases
and for understanding the emergence of dynamical order in complex quantum systems.

\begin{acknowledgments}
	This work is supported by National Key R\&D Program of China under Grant No.2022YFA1405800, the Key-Area Research and Development Program of Guangdong Province (Grant No.2018B030326001), Guangdong Provincial Key Laboratory(Grant No.2019B121203002).
\end{acknowledgments}

\section{End Matter}
\subsection{Proof of TLTC Vanishing in the Long-Time Limit} 

{\em Single-particle extended states.---} We show that for single-particle systems whose eigenstates are extended, the TLTC vanishes in the long-time limit. 
For extended eigenstates 
\(|\psi_n\rangle\) in a one-dimensional system, the local probability density exhibits a scaling behavior of
\(|\psi_n(j)|^2 \sim 1/L\)~\cite{AL4,Thouless}.  
The energy spectrum exhibits level repulsion, rendering the eigenenergies generally non-degenerate.
For an initial state \( |\psi(0)\rangle = |j_0\rangle \), 
and local observable \(O_j = |j\rangle\langle j|\), 
the local occupation probability at site \(j\) is 
\[
P_j(t) = |\langle j|\psi(t)\rangle|^2 
= \left| \sum_n c_n e^{-iE_n t} \psi_n(j) \right|^2.
\]
where $c_n = \langle\psi_n|j_0\rangle$. The TLTC reads
\begin{align*}\notag 
	\mathcal{C}_{a}^{(O)}(T) 
	&= \frac{1}{T}\int_0^T \big| P_{j_0}(t) - P_{j_0+a}(t) \big|^2 dt \\
	&= \sum_{m,n,p,q} c_m^* c_n c_p^* c_q 
	\left[ \frac{1}{T}\int_0^T e^{i(E_m - E_n + E_p - E_q)t} dt \right] \\
	&\quad \times \big[ \psi_m^*(j_0)\psi_n(j_0) - \psi_m^*(j_0+a)\psi_n(j_0+a) \big] \\
	&\quad \times \big[ \psi_p^*(j_0)\psi_q(j_0) - \psi_p^*(j_0+a)\psi_q(j_0+a) \big].
\end{align*}

As \(T \to \infty\), the oscillatory time integral selects terms satisfying
\(E_m - E_n + E_p - E_q = 0\), effectively producing
\(\delta_{E_m - E_n + E_p - E_q, 0}\).
The dominant contributions arise from two classes of index contractions.

\paragraph*{Case 1: \(m = n\), \(p = q\).}
\begin{align*}\notag 
	\mathcal{C}_{a}^{(O)}(T)= 
	\left(
	\sum_m |c_m|^2
	\big[ |\psi_m(j_0)|^2 - |\psi_m(j_0+a)|^2 \big]
	\right)^2.
\end{align*}
Extended states satisfy
$|\psi_m(j_0)|^2 - |\psi_m(j_0+a)|^2=O(1/L)$, hence this term scales as \(O(1/L^2)\).

\paragraph*{Case 2: \(m = q\), \(n = p\).}
\begin{align*}\notag 
	&\mathcal{C}_{a}^{(O)}(T)=\\
	&\sum_{m,n} |c_m|^2 |c_n|^2 
	\big|
	\psi_m^*(j_0)\psi_n(j_0)
	- \psi_m^*(j_0+a)\psi_n(j_0+a)
	\big|^2.
\end{align*}
For extended eigenstates, amplitudes at neighboring sites differ by \(O(1/L)\).
Cross terms with \(m \neq n\) average to zero due to rapid phase oscillations
from energy differences \(E_m - E_n \neq 0\).
Thus this contribution is \(O(1/L)\).

Both terms vanish as $L\rightarrow\infty$, proving that $\lim_{T\rightarrow\infty}\mathcal{C}_{a}^{(O)}(T)=0$ in the extended phase.

{\em Many-body ergodic phase.---}
For many-body systems obeying eigenstate thermalization hypothesis (ETH), 
the TLTC also vanishes. 
The proof relies on three standard properties of ergodic phases:

(i) Diagonal ensemble.  
In the long-time limit, the system approaches the diagonal ensemble
$\overline{\rho} = \sum_n |c_n|^2 |E_n\rangle\langle E_n|$,
where $\{|E_n\rangle\}$ are the energy eigenstates of $H$ and 
$c_n = \langle E_n|\psi(0)\rangle$.  
For any bounded operator $X$, the long-time average satisfies
\[
\lim_{T\rightarrow\infty}\frac{1}{T}\int_0^T 
\langle\psi(0)|X(t)|\psi(0)\rangle\,dt 
= \mathrm{Tr}(\overline{\rho}X).
\]
This follows from the fact that the eigenenergies $\{E_n\}$ are 
non-degenerate, so that the oscillatory terms $e^{i(E_m-E_n)t}$ vanish upon time averaging, 
leaving only the diagonal contributions with $E_m = E_n$.

(ii) ETH condition and self-averaging.  
ETH dictates that the expectation value of a local observable in an individual eigenstate closely approximates the microcanonical average~\cite{ETH1,ETH2,ETH3},
\[
\langle E_n | O_j | E_n \rangle \approx \langle O_j \rangle_\mathrm{micro}(E_n).
\]
In the thermodynamic limit, an ergodic system exhibits self-averaging: 
a single large realization becomes representative of the ensemble average.
As a result, spatial fluctuations of the microcanonical expectation values vanish,
\[
\langle O_j \rangle_\mathrm{micro}(E) \approx \langle O_{j+a} \rangle_\mathrm{micro}(E) 
\quad \text{for large } L.
\]
This emergent spatial homogeneity arises because extended eigenstates sample the entire system,
effectively averaging over the local disorder landscape.

Combining these two principles yields
\begin{align*}
&\langle E_n | O_j | E_n \rangle 
\approx \langle O_j \rangle_\mathrm{micro}(E_n) \\
&\approx \langle O_{j+a} \rangle_\mathrm{micro}(E_n) 
\approx \langle E_n | O_{j+a} | E_n \rangle,
\end{align*}
and consequently, in the diagonal ensemble,
\begin{align*}
&\mathrm{Tr}(\overline{\rho} O_j)
= \sum_n |c_n|^2 \langle E_n | O_j | E_n \rangle \\
&= \sum_n |c_n|^2 \langle E_n | O_{j+a} | E_n \rangle
= \mathrm{Tr}(\overline{\rho} O_{j+a}).
\end{align*}

(iii) Correlation decay.  
For bounded local operators $X$ and $Y$,  
two-time correlations decay rapidly such that
\[
\lim_{T\rightarrow\infty}\frac{1}{T}\int_0^T 
\langle\psi(0)|X(t)Y(t)|\psi(0)\rangle\,dt 
= \mathrm{Tr}(\overline{\rho}XY).
\]

Using properties (i)-(iii), 
we have
\begin{align*}
	\lim_{T\rightarrow\infty}\mathcal{C}_{a}^{(O)}(T)
	&= \lim_{T\rightarrow\infty}\frac{1}{T}\int_0^T 
	\langle\psi(0)|A(t) - \mathcal{T}_a[A(t)]|\psi(0)\rangle\\
	&\quad \times \langle\psi(0)|A(t) - \mathcal{T}_a[A(t)]|\psi(0)\rangle^* dt \\
	&= [\mathrm{Tr}(\overline{\rho}O_j)]^2 
	+ [\mathrm{Tr}(\overline{\rho}O_{j+a})]^2 
	- 2[\mathrm{Tr}(\overline{\rho}O_j)]^2 \\
	&= 0,
\end{align*}
where $A(t) = U^\dagger(t)O_jU(t)$. The vanishing of the TLTC in ergodic phases reflects a key feature of thermalization: long-time dynamics restores translational symmetry even when the Hamiltonian is not translationally invariant. 

\begin{thebibliography}{36}
\bibitem{explainopen} For example, in open quantum systems, the characterization of topological properties remains a subject of debate. The present definition of the TLTC can be directly applied to open systems, providing a unified framework that allows consistent characterization of topology in both closed and open settings.
\end{thebibliography}
\end{document}